\documentclass[aps,prl,preprint,superscriptaddress,showpacs]{revtex4}
\usepackage{epsfig}

\newcommand{\be}{\begin{equation}}
\newcommand{\ee}{\end{equation}}
\newcommand{\ba}{\begin{eqnarray}}
\newcommand{\ea}{\end{eqnarray}}
\newcommand{\bd}{\begin{displaymath}}
\newcommand{\ed}{\end{displaymath}}

\def\fourth{{\textstyle{\frac{4}{3}}}}

\begin{document}

\title{ On the Strongly-Interacting Low-Viscosity Matter Created in Relativistic 
Nuclear Collisions }

\author{Laszlo P. Csernai} 
\affiliation{Section for Theoretical Physics, Department of Physics,
University of Bergen, Allegaten 55, 5007 Bergen, Norway}
\affiliation{ MTA-KFKI, Research Institute of Particle and Nuclear Physics,
1525 Budapest 114, P. O. Box 49, Hungary}
\author{Joseph I. Kapusta}
\affiliation{School of Physics and Astronomy, University of Minnesota,
Minneapolis, MN 55455, USA}
\author{Larry D. McLerran}
\affiliation{Nuclear Theory Group and Riken Brookhaven Center, Brookhaven 
National Laboratory, Bldg. 510A, Upton, NY 11973}

\date{6 September 2006}

\begin{abstract}

Substantial collective flow is observed in collisions between large nuclei at 
RHIC (Relativistic Heavy Ion Collider) as evidenced by single-particle 
transverse momentum distributions and by azimuthal correlations among 
the produced particles.  The data are well-reproduced by perfect fluid dynamics.  
A calculation of the dimensionless ratio of shear viscosity $\eta$ to entropy 
density $s$ by Kovtun, Son and Starinets within anti-de Sitter space/conformal 
field theory yields $\eta/s = \hbar/4\pi k_B$ which has been conjectured to be a 
lower bound for any physical system.  Motivated by these results, we show that 
the transition from hadrons to quarks and gluons has behavior similar to helium, 
nitrogen, and water at and near their phase transitions in the ratio $\eta/s$. 
We suggest that experimental measurements can pinpoint the location of this 
transition or rapid crossover in QCD.  

\end{abstract}

\vspace*{0.5in}


\maketitle

One of the amazing experimental discoveries of measurements on gold-gold 
collisions at RHIC (Relativistic Heavy Ion Collider) at Brookhaven National 
Laboratory is the surprising amount of collective flow exhibited by the outgoing 
hadrons.  Collective flow is evidenced in both the single-particle transverse 
momentum distribution \cite{radial}, commonly referred to as radial flow, and in 
the asymmetric azimuthal distribution around the beam axis \cite{RHICv2}, 
quantified by the functions $v_1(y,p_T), v_2(y,p_T), ...$ in the expansion
\be
\frac{d^3N}{dydp_Td\phi} = \frac{1}{2\pi}\frac{d^2N}{dydp_T} \left[ 1 
+ 2v_1(y,p_T) \cos(\phi) + 2v_2(y,p_T) \cos(2\phi) + \cdot\cdot\cdot \right]
\ee
where $y$ is the rapidity and $p_T$ is the transverse momentum.  The function 
$v_2(y=0,p_T)$, in particular, was expected to be much smaller at RHIC than it 
is at the lower energies of the SPS (Super Proton Synchrotron) at CERN 
\cite{SPSv2}, but in fact it is about twice as large.  Various theoretical 
calculations \cite{coalesce} support the notion that collective flow is mostly 
generated early in the nucleus-nucleus collision, and is present at the partonic 
level before partons coalesce or fragment into hadrons.  Theoretical 
calculations including only two-body interactions between partons cannot 
generate sufficient flow to explain the observations unless partonic cross 
sections are artificially enhanced by more than an order of magnitude over 
perturbative QCD predictions \cite{Molnar}.  This has emphasized that the 
quark-gluon matter created in these collisions is strongly interacting, unlike 
the type of weakly interacting quark-gluon plasma expected to occur at very high 
temperatures on the basis of asymptotic freedom \cite{pQCD}.  On the other hand, 
lattice QCD calculations yield an equation of state that differs from an ideal 
gas only by about 10\% once the temperature exceeds $1.5 T_c$, where $T_c 
\approx 175$ MeV is the critical or crossover temperature from quarks and gluons 
to hadrons \cite{latticeQCD}.  Furthermore, perfect fluid dynamics (with zero 
shear and bulk viscosities) reproduces the measurements of radial flow and $v_2$ 
very well up to transverse momenta of order 1.5 GeV/c \cite{hydro}.

An amazing theoretical discovery was made by Kovtun, Son and Starinets 
\cite{Son}, who showed that certain special field theories, special in the sense 
that they are dual to black branes in higher space-time dimensions, have the 
ratio $\eta/s = 1/4\pi$ (we use units with $\hbar = k_{\rm B} = c =1$) where 
$\eta$ is the shear viscosity and $s$ is the entropy density.  They conjectured 
that all substances have this value as a lower limit, and gave as examples 
helium, nitrogen, and water at pressures of 0.1 MPa, 10 MPa, and 100 MPa, 
respectively.  Interesting enough, this bound is also obeyed by ${\cal N}=4$ 
supersymmetric $SU(N_c)$ Yang-Mills theory in the large $N_c$ limit \cite{SUSY}.  

We are motivated by these discoveries to study what happens in QCD at finite 
temperature.  The relatively good agreement between perfect fluid calculations 
and experimental data for hadrons of low to medium transverse momentum at RHIC 
suggests that the viscosity is small; however, it cannot be zero.  Indeed, the 
calculations within anti-de Sitter space/conformal field theory suggests that 
$\eta \ge s/(4\pi)$.  Our conclusion will be that sufficiently precise 
calculations and measurements should allow for a determination of the ratio 
$\eta/s$ as a function of temperature, and that this ratio can pinpoint the 
location of the phase transition or rapid crossover from hadronic to quark and 
gluon matter.  This is a different method than trying to infer the equation of 
state of QCD in the form of pressure $P$ as a function of temperature $T$ or 
energy density $\epsilon$.

The energy-momentum tensor density for a perfect fluid (which does not imply 
that the matter is non-interacting) is $T^{\mu\nu} = -
Pg^{\mu\nu}+wu^{\mu}u^{\nu}$.  Here $w=P+\epsilon=Ts$ is the local enthalpy 
density and $u^{\mu}$ is the local flow velocity.  Corrections to this 
expression are proportional to first derivatives of the local quantities whose 
coefficients are the shear viscosity $\eta$ and bulk viscosity $\zeta$.  
(Thermal conductivity is neither relevant nor defined when all net conserved 
charges, such as electric charge and baryon number, are zero.)  Explicit 
expressions may be found in textbooks \cite{fluid}.  Perfect fluid dynamics 
applies when the viscosities are small, or when the gradients are small, or 
both.  The dispersion relations for the transverse and longitudinal (pressure) 
parts of the momentum density are
\ba
\omega + i D_t k^2 &=& 0 \nonumber \\
\omega^2 - v^2 k^2 + i D_l \omega k^2 &=& 0
\ea
where $D_t = \eta/w$ and $D_l = (\fourth \eta + \zeta)/w$ are diffusion 
constants with the dimension of length and $v$ is the speed of sound.  Since 
$w=Ts$, and since usually the bulk viscosity is small compared to the shear 
viscosity, the dimensionless ratio of (shear) viscosity to entropy (disorder) 
$\eta/s$ is a good way to characterize the intrinsic ability of a substance to 
relax towards equilibrium independent of the actual physical conditions 
(gradients of pressure, energy density, etc.).  It is also a good way to compare 
very different substances.

In figures 1 through 3 we plot the ratio $\eta/s$ versus temperature at three 
fixed pressures, one of them being the critical pressure (meaning that the curve 
passes through the critical point) and the other ones being larger and smaller, 
for helium, nitrogen and water.  The ratio was constructed with data obtained 
from the National Institute of Standards and Technology (NIST) \cite{NIST}.  
(Care must be taken to absolutely normalize the entropy to zero at zero 
temperature; we did that using data from CODATA \cite{CODATA}.)  The important 
observation \cite{private,singular} is that $\eta/s$ has a minimum at the 
critical point where there is a cusp.  At pressures below the critical pressure 
there is a discontinuity in $\eta/s$, and at pressures above it there is a broad 
smooth minimum. The simplest way to understand the general behavior was 
presented by Enskog, as explained in \cite{atomtheory}.  Shear viscosity 
represents the ability to transport momentum.  In classical transport theory of 
gases $\eta/s \sim T l_{\rm free} \bar{v}$, where $l_{\rm free}$ is the mean 
free path and $\bar{v}$ is the mean speed.  For a dilute gas the mean free path 
is large, $l_{\rm free} \sim 1/n\sigma$, with $n$ the particle number density 
and $\sigma$ the cross section.  Hence it is easy for a particle to carry 
momentum over great 
distances, leading to a large viscosity.  (This is the usual paradox, that a 
nearly ideal classical gas has a divergent viscosity.)  In a liquid there are 
strong correlations between neighboring atoms or molecules.  A liquid is 
homogeneous on a mesoscopic scale, but on a microscopic scale it is a mixture of 
clusters and voids.  The action of pushing on one atom is translated to the next 
one and so on until a whole row of atoms moves to fill a void, thereby 
transporting momentum over a relatively large distance and producing a large 
viscosity.  Reducing the temperature at fixed pressure reduces the density of 
voids, thereby increasing the viscosity.  The viscosity, normalized to the 
entropy, is observed to be the smallest at or near the critical temperature, 
corresponding to the most difficult condition to transport momentum.  This is an 
empirical observation.  The mean free path must lie somewhere between the dilute 
gas limit, $1/n\sigma$, and the close-packing limit, $1/n^{1/3}$.  For a 
massless gas of $N$ bosonic degrees of freedom, with entropy density $s=N 
(4\pi^2/90) T^3$, the close-packed limit gives $\eta/s \approx 2/N^{1/3}$.

How does this relate to hadrons and quark-gluon plasma?  In the low energy 
chiral limit for pions the cross section is proportional to $\hat{s}/f_{\pi}^4$, 
where $\hat{s}$ is the usual Mandelstam variable for invariant mass-squared and 
$f_{\pi}$ is the pion decay constant.  The thermally averaged cross section is 
$\langle \sigma \rangle \propto T^2/f_{\pi}^4$, which leads to $\eta/s \propto 
(f_{\pi}/T)^4$.  Explicit calculation gives \cite{Prakash}
\be
\frac{\eta}{s} = \frac{15}{16\pi} \frac{f_{\pi}^4}{T^4}
\label{chiral}
\ee
Thus the ratio $\eta/s$ diverges as $T \rightarrow 0$.  At the other extreme 
lies quark-gluon plasma.  The parton cross section behaves as $\sigma \propto 
g^4/\hat{s}$.  A first estimate yields $\eta/s \propto 1/g^4$.  Asymptotic 
freedom at one loop order gives $g^2 \propto 1/\ln(T/\Lambda_T)$ where 
$\Lambda_T$ is proportional to the scale parameter $\Lambda_{\rm QCD}$ of QCD.  
Therefore $\eta/s$ is an increasing function of $T$ in the quark-gluon phase.  
As a consequence, $\eta/s$ must have a minimum.  Based on atomic and molecular 
data, this minimum should lie at the critical temperature if there is one, 
otherwise at or near the rapid crossover temperature.

The most accurate and detailed calculation of the viscosity in the low 
temperature hadron phase was performed in \cite{Prakash}.  The two-body 
interactions used went beyond the chiral approximation, and included 
intermediate resonances such as the $\rho$-meson.  The results are displayed in 
figure 4, both two flavors (no kaons) and three flavors (with kaons).  The 
qualitative behavior is the same as in eq. (\ref{chiral}).  The most accurate 
and detailed calculation of the viscosity in the high temperature quark-gluon 
phase was performed in \cite{Arnold}.  They used perturbative QCD to calculate 
the full leading-order expression, including summation of the Coulomb 
logarithms.  For three flavors of massless quarks the result is
\be
\frac{\eta}{s} = \frac{5.12}{g^4 \ln(2.42/g)}
\ee
We used this together with the two-loop renormalization group expression for the 
running coupling
\be
\frac{1}{g^2(T)} = \frac{9}{8\pi^2} \ln\left( \frac{T}{\Lambda_T} \right)
+ \frac{4}{9\pi^2} \ln \left( 2 \ln\left( \frac{T}{\Lambda_T} \right) \right)
\ee
with $\Lambda_T = 30$ MeV, which approximately corresponds to using an energy 
scale of $2\pi T$ and $\Lambda_{\overline{MS}} = 200$ MeV.  The result is also 
plotted in figure 4.  These results imply a minimum in the neighborhood of the 
expected value of $T_c \approx 190$ MeV.  Whether there is a discontinuity or a 
smooth crossover cannot be decided since both calculations are unreliable near 
$T_c$.

It is interesting to ask what happens in the large $N_c$ limit with $g^2N_c$ 
held fixed \cite{largeN}.  In this limit, meson masses do not change very much 
but baryon masses scale proportional to $N_c$; therefore, baryons may be 
neglected in comparison to mesons due to the Boltzmann factor.  Since the meson 
spectrum is essentially unchanged with increasing $N_c$, so is the Hagedorn 
temperature.  The critical temperature to go from hadrons to quarks and gluons 
is very close to the Hagedorn temperature, so that $T_c$ is not expected to 
change very much either.  In the large $N_c$ limit the meson-meson cross section 
scales as $1/N_c^2$.  According to our earlier discussion on the classical 
theory of gases, this implies that the ratio $\eta/s$ in the hadronic phase 
scales as $N_c^2$.  This general result is obeyed by (\ref{chiral}) since it is 
known that $f_{\pi}^2$ scales as $N_c$.  The large $N_c$ limit of the viscosity 
in the quark and gluon phase may be inferred from the calculations of 
\cite{Arnold} to be
\be
\left(\frac{\eta}{s}\right)_{\rm QGP} = 
\left( \frac{1+3.974r}{1+1.75r} \right)
\frac{69.2}{\left(g^2N_c\right)^2 \ln \left( 26/(g^2N_c(1+0.5r))\right)}
\ee
where $r = N_f/N_c$.  Thus the ratio $\eta/s$ has a finite large $N_c$ limit in 
the quark and gluon phase.  Therefore, we conclude that $\eta/s$ has a 
discontinuity proportional to $N_c^2$ if $N_c \rightarrow \infty$.  This jump is 
in the opposite direction to that in figure 4.

So far the only quantitative results for viscosity in lattice gauge theory have 
been reported by Nakamura and Sakai \cite{latticeeta} for pure SU(3) without 
quarks.  This bold effort obtained $\eta/s \approx 1/2$ in the temperature range 
$1.6 < T/T_c < 2.2$, albeit with uncertainties of order 100\%.  Gelman, Shuryak 
and Zahed \cite{cQCD} have modeled the dynamics of long wavelength modes of QCD 
at temperatures from $T_c$ to $1.5 T_c$ as a classical, nonrelativistic gas of 
massive quasi-particles with color charges.  They obtained a ratio of $\eta/s 
\approx 0.34$ in this temperature range.

It ought be possible to extract numerical values of the viscosity in heavy ion 
collisions via scaling violations to perfect fluid flow predictions.  One should 
perform a systematic beam energy and projectile/target mass scan from SPS 
energies to the top RHIC energy, and then on to the LHC. Flow data, in the form 
of the functions $v_1, v_2, ... $, should be obtained and compared with the 
results of calculations based on relativistic viscous fluid dynamics.  This 
program is analogous to what was accomplished at lower energies of 30 to 1000 
MeV per nucleon beam energies in the lab frame.  At those energies, scaling 
violations to perfect fluid dynamics were indeed observed \cite{Bonasera}.  It 
was possible to infer the compressibility of nuclear matter and the
momentum-dependence of the nuclear optical potential via the transverse momentum 
distribution relative to the reaction plane \cite{BUU} and via the balance 
between attractive and repulsive scattering \cite{balance}.  There is much to be 
learned about QCD at high energy densities.

We thank D. T. Son for helpful discussions.  This work was supported by the US 
Department of Energy under grants DE-FG02-87ER40328 (J. Kapusta) and DE-AC02-
98CH10886 (L. McLerran).

\begin{figure}
 \centering
 \includegraphics[width=3.5in,angle=90]{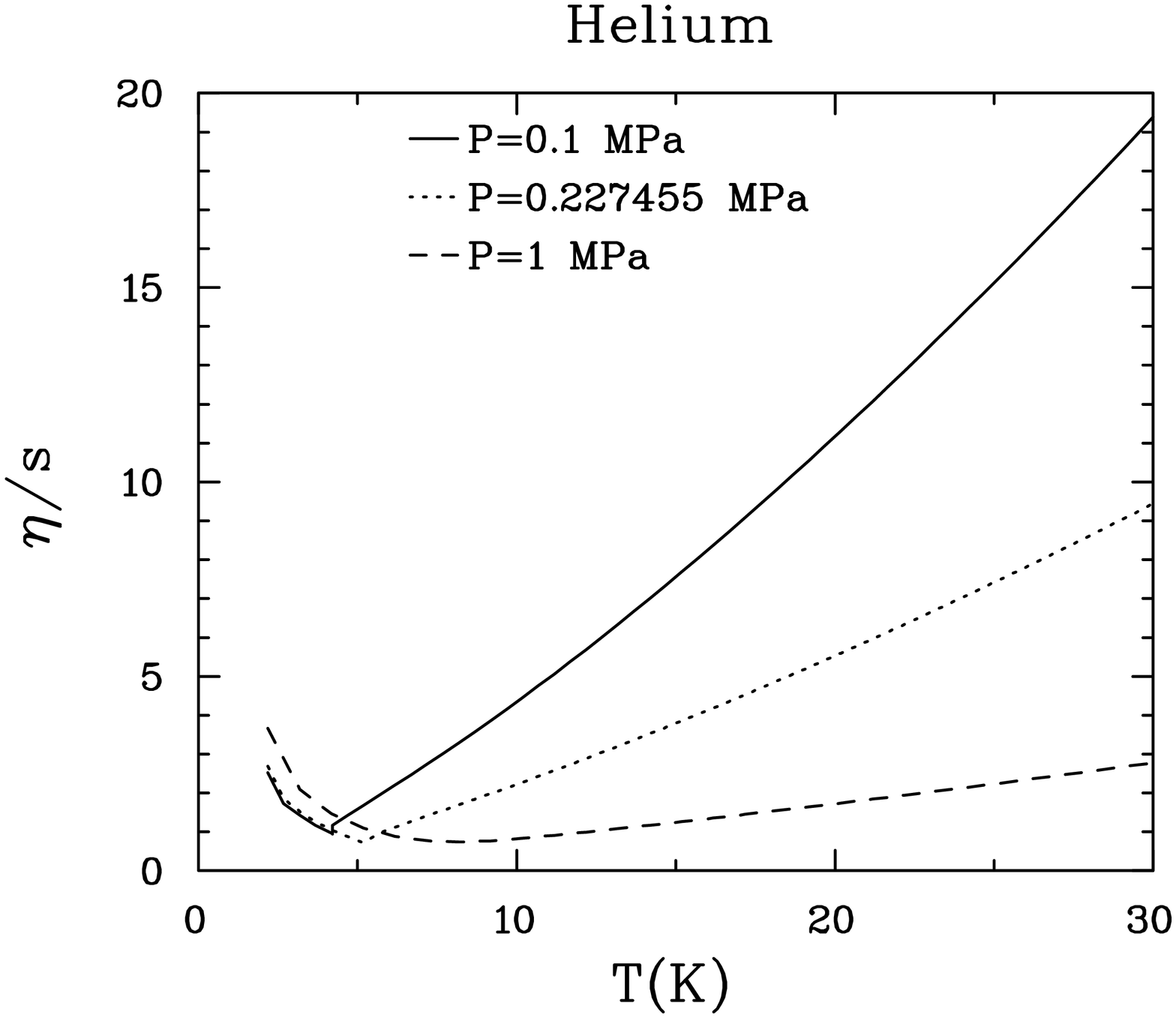}
 \caption{The ratio $\eta/s$ as a function of $T$ for helium with $s$ normalized 
such that $s(T=0)=0$.  The curves correspond to fixed pressures, one of them 
being the critical pressure, and the others being greater (1 MPa) and the other 
smaller (0.1 MPa).  Below the critical pressure there is a jump in the ratio, 
and above the critical pressure there is only a broad minimum. They were 
constructed using data from NIST and CODATA.}
 \label{fig1}
\end{figure}

\begin{figure}
 \centering
 \includegraphics[width=3.5in,angle=90]{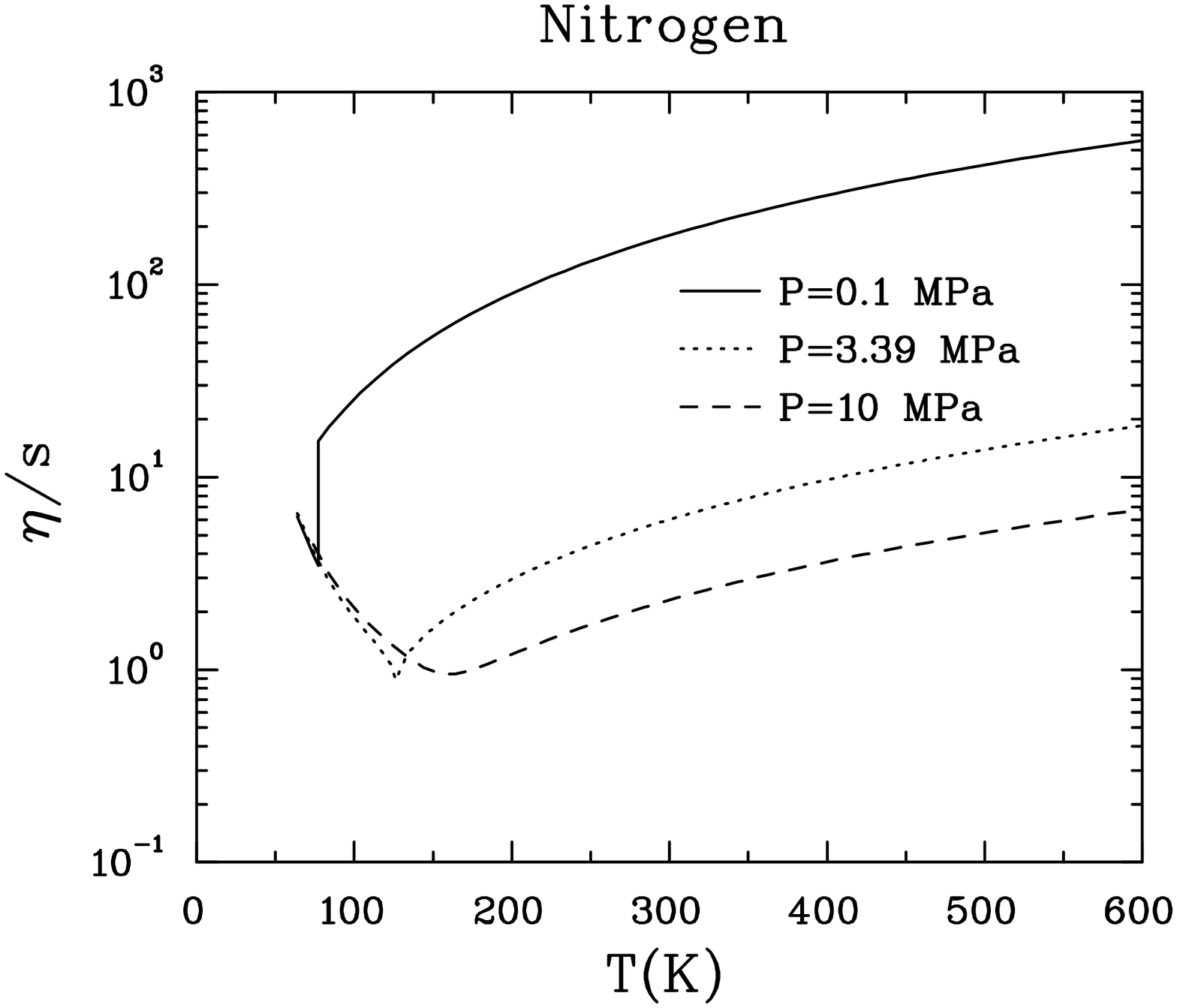}
 \caption{The ratio $\eta/s$ as a function of $T$ for nitrogen with $s$ 
normalized such that $s(T=0)=0$.  The curves correspond to fixed pressures, one 
of them being the critical pressure, and the others being greater (10 MPa) and 
the other smaller (0.1 MPa).  Below the critical pressure there is a jump in the 
ratio, and above the critical pressure there is only a broad minimum. They were 
constructed using data from NIST and CODATA.  The curves are plotted on 
logarithmic scale to make the behavior around the critical point more visible.}
 \label{fig2}
\end{figure}

\begin{figure}
 \centering
 \includegraphics[width=3.5in,angle=90]{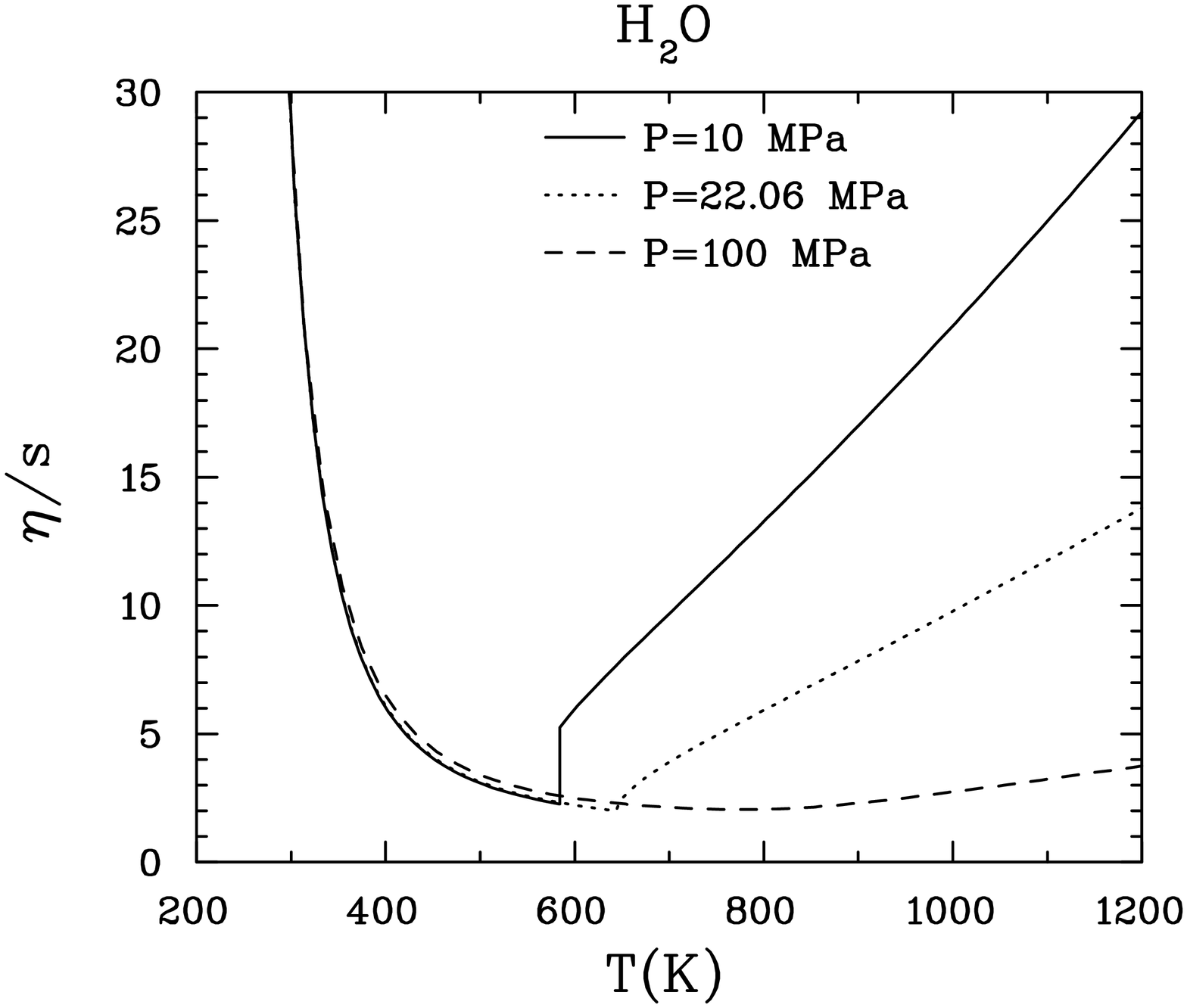}
 \caption{ The ratio $\eta/s$ as a function of $T$ for water with $s$ normalized 
such that $s(T=0)=0$.  The curves correspond to fixed pressures, one of them 
being the critical pressure, and the others being greater (100 MPa) and the 
other smaller (10 MPa).  Below the critical pressure there is a jump in the 
ratio, and above the critical pressure there is only a broad minimum. They were 
constructed using data from NIST and CODATA.}
 \label{fig3}
\end{figure}

\begin{figure}
 \centering
 \includegraphics[width=3.5in,angle=90]{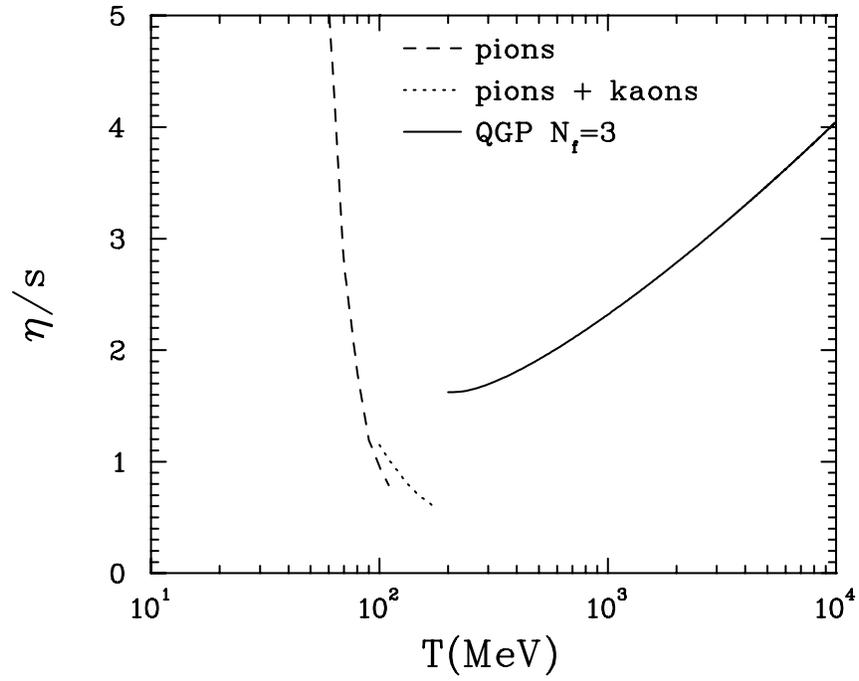}
 \caption{The ratio $\eta/s$ for the low temperature hadronic phase and for the 
high temperature quark-gluon phase.  Neither calculation is very reliable in the 
vicinity of the critical or rapid crossover temperature.}
 \label{fig4}
\end{figure}

\end{document}